\documentclass[pra,letterpaper,twocolumn,showpacs,superscriptaddress,floatfix]{revtex4}

\usepackage{graphicx}

\usepackage{dcolumn}% Align table columns on decimal point
\usepackage{bm}% bold math

\begin{document}

\title{Ehrenfest approach to open double-well dynamics}

\author{Stephen Choi}

\affiliation{Department of Physics, University of Massachusetts, Boston, MA 02125, USA}

\author{Roberto Onofrio}

\affiliation{Dipartimento di Fisica e Astronomia ``Galileo Galilei'', Universit\`a  di Padova, \\
Via Marzolo 8, Padova 35131, Italy}

\affiliation{Department of Physics and Astronomy, Dartmouth College, 6127 Wilder Laboratory, 
Hanover, NH 03755, USA}

\author{Bala Sundaram}

\affiliation{Department of Physics, University of Massachusetts, Boston, MA 02125, USA}

\begin{abstract}
We consider an Ehrenfest approximation for a particle in a double-well potential in the 
presence of an external environment schematized as a finite resource heat bath. 
This allows us to explore how the limitations in the applicability of Ehrenfest 
dynamics to nonlinear systems are modified in an open system setting.  
Within this framework, we have identified an environment-induced spontaneous symmetry 
breaking mechanism, and we argue that the Ehrenfest approximation becomes increasingly 
valid in the limit of strong coupling to the external reservoir, either in the form of 
increasing number of oscillators or increasing temperature. 
The analysis also suggests a rather intuitive picture for the general phenomenon of quantum 
tunneling and its interplay with classical thermal activation processes, which may be 
of relevance in physical chemistry, ultracold atom physics, and fast-switching dynamics such 
as in superconducting digital electronics.
\end{abstract}


\maketitle

\section{Introduction}

The analysis of the relationship between full quantum solutions and their semiclassical 
or classical counterparts has played an important role in modern physics both for promoting
a better conceptual understanding of its foundations and for symplifying complex problems. 
Prominent among the many techniques introduced to treat quantum systems is the approximation 
known as Ehrenfest dynamics~\cite{Ehrenfest} extensively used both for its generality 
and ready comparison to the classical limit. This latter feature has resulted in its use 
in the analysis of nonlinear dynamical systems, including those which exhibit chaotic 
properties~\cite{Chaos}. 

Here we consider a one-dimensional dynamical system with nontrivial properties describing 
a particle moving under the simultaneous influence of a double-well potential and a heat 
bath defined in terms of its density of states and temperature. 
The case of a double-well is very important for various reasons, and an Ehrenfest treatment 
is obviously going to be approximate giving rise to misleading results on long timescales 
as discussed in~\cite{PS1,PS2,PS3,SM}. Nevertheless this is a study of intrinsic value for 
various reasons. 

First, the double-well is a paradigmatic example of a dynamical system sufficiently different from 
that of the harmonic oscillator, with a rich interplay between intrawell and interwell dynamics. 
Second, the shortcomings of Ehrenfest dynamics may be mitigated in contexts where the dynamics 
is frequently switched.  Superconducting digital electronics provides an example for which the 
dynamics is frequently reset and, as such, no significant traces of the deviations between the 
Ehrenfest and the exact, quantum, dynamics should appear even on long time scales. 
Further, the double-well description here allows naturally for defining a dichotomous 
variable associated with quantities which are functions of the two potential minima.  
Third, there is a crossover regime between classical and quantum stochastic resonance in double-well 
systems \cite{Gammaitoni}, and the Ehrenfest dynamics may help fill this gap with simple, easily recognizable, 
dynamical structures, a sort of ``coarse graining'' of the full Hilbert space. 
In particular, we are able to interpolate between the situation of a purely quantum closed system 
dynamics, with possibility of tunneling, and an open semiclassical or classical system in which 
hopping between the two minima of the double-well potential is instead achieved via thermal 
activation. In the open system case, the reservoir is schematized as a number of harmonic 
oscillators, with uniform density of frequencies in a finite bandwidth, linearly coupled to 
the particle in the double-well. 
This represents the closest approximant to the ideal case of the Caldeira-Leggett model with infinite 
oscillators~\cite{Caldeira,CaldeiraRMP}, and can also be considered as a general model for describing 
solid-state devices. 

The paper is structured as follows. In Section II we introduce the model and discuss 
a feature which emerges already at the classical level, a sort of phase transition from 
a bistable to a monostable potential with increasing number of bath oscillators.
In Section III we discuss the Ehrenfest approximation to the double-well potential 
and introduce the extended phase space structure. 
Numerical results for the Ehrenfest model of a double-well are reported in Section IV, 
including the classical limit and the Fourier analysis of the particle motion.
In the concluding section we discuss the relevance of our considerations for the debate 
surrounding the Ehrenfest construction as applied to the double-well model, arguing that 
in the macroscopic limit, {\it i.e.} with large number of oscillators, and/or in 
the high-temperature regime, the Ehrenfest approximation becomes more valid. 

\section{Classical considerations}

The double-well external potential acting on a particle of mass $M$ is assumed to be of the form 
\begin{equation}
V(Q) =  -\mu Q^2 +\lambda Q^4,
\end{equation}
where the two minima of the potential occur at $\pm \sqrt{\mu/(2\lambda)}$, and an 
energy barrier, relevant to the interwell dynamics, equal to $\Delta E=-\mu^2/(4\lambda)$. 
In the presence of a translationally invariant heat bath, the total Hamiltonian becomes

\begin{equation}
H_{tot} = \frac{P^2}{2M} -\mu  Q^2 + \lambda Q^4 + \sum_{n=1}^N 
\left[\frac{p_{n}^2}{2m} + \frac{1}{2}m \omega_{n}^2 (q_n-Q)^2  \right ],
\end{equation}
which can be easily regrouped as
\begin{eqnarray}
H_{tot} & = & \frac{P^2}{2M} - \left(\mu-\frac{m}{2} \sum_{n=1}^N \omega_n^2\right) Q^2 + \lambda Q^4 - 
\nonumber \\
& & m (\sum_{n=1}^N \omega_n^2 q_n) Q + \sum_{n=1}^N \left( \frac{p_n^2}{2m} + 
\frac{1}{2}m \omega_n^2 q_n^2 \right).
\end{eqnarray}

\begin{figure}[t] 
\begin{center}
{\includegraphics[width=\columnwidth]{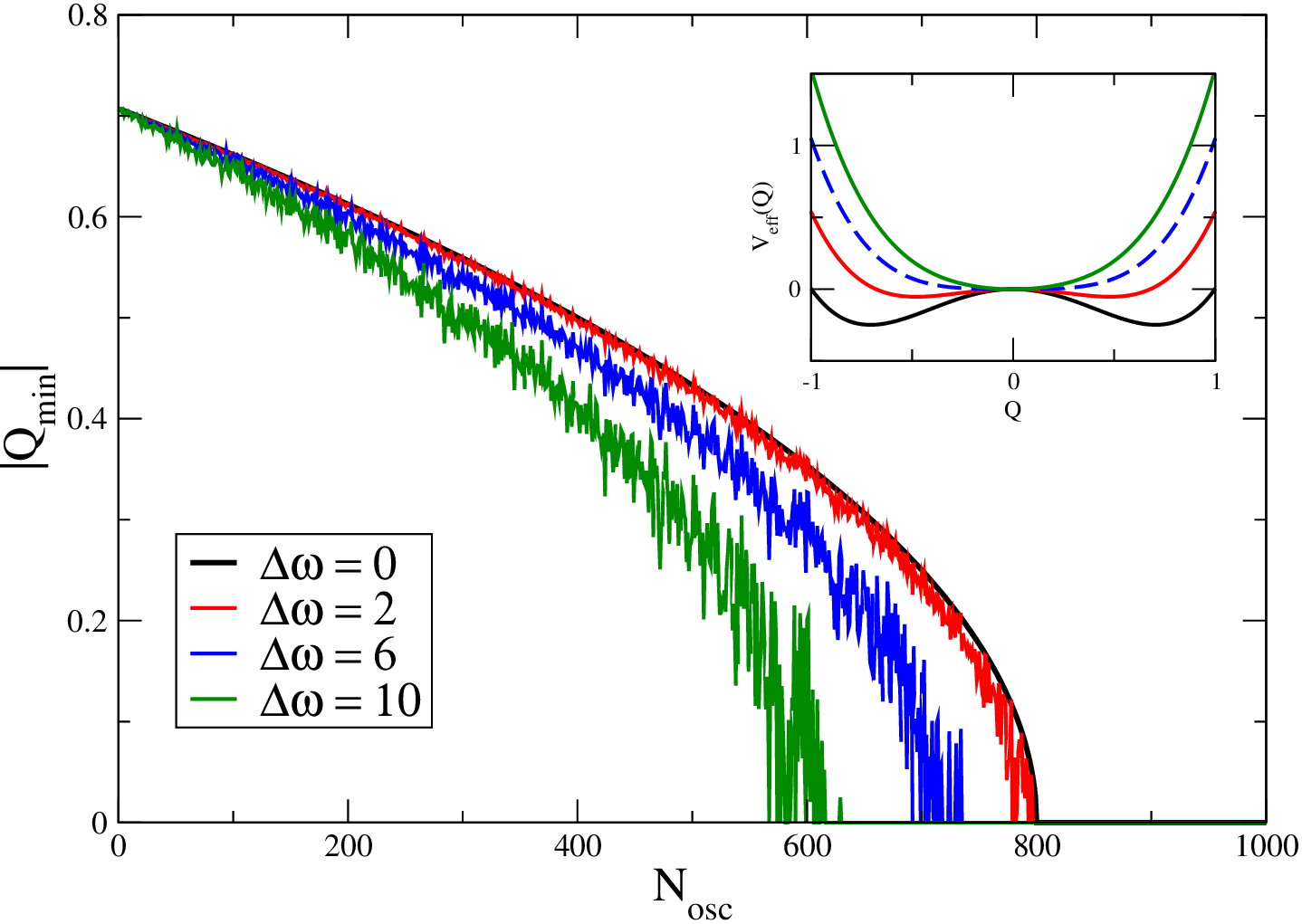}}
\caption{(Color online) Transition from bistable to monostable behavior induced by the 
external environment. 
The absolute value of the minimum of the potential is plotted versus the number of involved 
harmonic oscillators with uniform distribution of their angular frequency in a range centered 
around $\omega_0=5$, for different values of their bandwidth $\Delta \omega=0,2,6,10$ (from the top to 
the bottom curves, black, red, blue, green respectively). The case of monochromatic bath yields the 
only deterministic curve and gives a value of $N_{\mathrm{crit}}=2 \mu/(m \omega_0^2)=800$ 
for $\mu=1$ and $m=10^{-4}$, both in arbitrary units. 
The opposite case of maximal bandwidth yields instead a value which is smaller by a 
factor $1+\Delta \omega^2/(12\omega_0^2)=4/3$, {\it i.e.} $N_{\mathrm{crit}}=600$, in 
good agreement with the observed behavior of the curve at the endpoint. 
The inset explicitly shows the changing effective potential for increasing number of oscillators, 
$N_{\mathrm{osc}}=1,300,600,900$ from the bottom to the top curves (black, red, blue, green respectively), 
with the dashed curve indicating the threshold case of $\langle N_{\mathrm{crit}} \rangle=600$. 
The bandwidth in this case was taken to the maximum value, $\Delta \omega=10$.}
\label{Josephsonfig1}
\end{center}
\end{figure}

Notice that in the large $N$ limit, if the initial positions of all particles are all 
randomly distributed (with their sign too), the linear term in Q is negligible. 
In the realistic case of finite $N$, this term will induce an external fluctuating, 
spatially-independent, force on the test particle resulting from the finite 
number of kicks due to the bath particles. 
Also, the effect of the heat bath is a renormalization of the quadratic term of the 
test particle potential energy, i.e. $\mu \mapsto \mu- m\langle \omega^2 \rangle N/2$ 
where $\langle \omega^2 \rangle=\sum_{n=1}^N \omega_n^2/N$.
This is important because the heat bath renormalization term may change qualitatively the 
behavior which, in the large $N$ limit, will end up in a quadratic plus quartic oscillator 
with no bistability. The average critical number of oscillators which defines the transition 
from a bistable to a monostable potential is easily written as 
$\langle N_{\mathrm{crit}} \rangle=2 \mu/(m \langle \omega^2 \rangle)$. 
By assuming a uniform distribution of the frequencies in the interval 
$[\omega_{\mathrm{min}},\omega_{\mathrm{max}}]$, we obtain 
\begin{eqnarray}
\langle N_{\mathrm{crit}}& \rangle=&6\mu/[m(\omega_{\mathrm{max}}^2+
\omega_{\mathrm{max}}\omega_{\mathrm{min}}+\omega_{\mathrm{min}}^2)]= \nonumber \\
& & \frac{2 \mu}{m \omega_0^2} \frac{1}{1+\Delta \omega^2/(12 \omega_0^2)},
\end{eqnarray}
where in the last expression we have introduced the average angular frequency
$\omega_0=(\omega_{\mathrm{max}}+\omega_{\mathrm{min}})/2$ and the bandwidth 
$\Delta \omega=\omega_{\mathrm{max}}-\omega_{\mathrm{min}}$.  
In Fig.~\ref{Josephsonfig1} we plot the absolute value of the minimum of the double-well 
potential, an indicator of the transition from bistable to monostable behavior, versus 
the number of oscillators. The presence of an increasing bath frequency bandwidth decreases 
the critical value of the number of oscillators for the transition, and the fact that 
these frequencies are randomly selected results in a stochastic component to the potential 
as well as in the location of its minimum. This provides an example of spontaneous symmetry breaking in which the environment restores the full symmetry of the ground state when the number of harmonic oscillators effectively interacting 
with the particle exceeds a threshold. The situation discussed here is similar to the one discussed in Ref. \cite{Jona}, where localization is induced in a gas of pyramidal molecules (for which the interatomic 
interaction can be also approximated by a double-well potential) by the presence of the other molecules. 
Both situations can be considered cases in which there is a density-dependent phase transition. 
Notice that once the double-well potential is introduced, regardless of its microscopic origin, the 
phase transition occurs already in the classical limit, as our discussion thus far does not involve 
any consideration of quantum effects.

\section{Ehrenfest approach to the double-well potential}

In order to analyze the quantum case within the Ehrenfest framework, we introduce 
the Heisenberg equations associated with the closed system Hamiltonian 
$\hat{H} = \hat{P}^2/2M + V(\hat{Q},t)$
\begin{equation}
\frac{d\hat{Q}}{dt} =  \frac{\hat{P}}{M} \;,\hspace{0.5cm}
\frac{d\hat{P}}{dt} = - \frac{\partial V(\hat{Q},t)}{\partial Q} \;.
\end{equation}
Each operator can be written as $\hat{A} =  \langle \hat{A} \rangle +\Delta
\hat{A}$, where $\langle \ldots \rangle$ denotes the expectation value so that
$\langle \Delta \hat{A} \rangle =0$.  Taylor expanding the
potential $V(\hat{Q},t)$ about $\langle \hat{Q} \rangle$ leads to 
the Ehrenfest equations 
\begin{eqnarray}
\frac{d \langle \hat{Q} \rangle}{dt} &=& \frac{\langle \hat{P} \rangle}{M} \;,  \label{Ehrenfest1} \\
\frac{d \langle \hat{P} \rangle}{dt} &=& - \sum_{n=0}^{\infty}
\frac{1}{n!} V^{(n+1)} (\langle \hat{Q} \rangle) \langle \Delta \hat{Q}^n \rangle \;,
\end{eqnarray}
where $V^{(n)} = \partial^{n} V/\partial Q^{n}$. Writing down the
corresponding evolution equations for $\langle \Delta\hat{Q}^n\rangle$ 
leads to an infinite hierarchy of equations \cite{PS1,PS2,PS3,SM}. While we summarize 
below the formalism with the goal of applying it to a double-well potential, we refer the 
interested reader to our recent discussion in Section II of \cite{Choionba} for a detailed 
description of the delicate interplay between the Heisenberg equations, the infinite hierarchy 
of Ehrenfest equations, and their  truncation by using the Gaussian approximation.

Due to the presence of a quartic term in the polynomial form of the potential energy for a 
double-well, moments across different orders are coupled to each other, and  higher-order 
moments grow and eventually become significant even if they were initially zero. 

The Ehrenfest expansion in the Gaussian approximation becomes  
\begin{eqnarray}
\frac{d  Q }{dt}  & = &  \frac{P}{M}  \;,  \\
\frac{d  P }{dt}  &  = & - \sum_{n = 0}^{\infty} V^{(2n+1)}(Q) \frac{\rho^{2n}}{n! 2^{n}} \;, \\
\frac{d \rho}{dt}  & = & \frac{\Pi}{M}  \;, \\
\frac{d \Pi}{dt}  & = & \frac{\hbar^2}{4 M \rho^3} -  
\sum_{n =0}^{\infty}  V^{(2n+2)}(Q) \frac{\rho^{2n+1}}{n! 2^n} \;,
\end{eqnarray}
where $Q \equiv \langle \hat{Q} \rangle$ and $P \equiv \langle \hat{P}
\rangle$ are the expectation values of position and momentum, respectively. 
Here, odd cumulants are identically zero and even cumulants can be
written in terms of variable $\rho$ as $\langle \Delta \hat{Q}^{2n} \rangle= 
\rho^{2n}/((2n!)2^{n}n!)$. As done earlier~\cite{PS1,PS2,PS3}, we also introduce a new variable 
$\Pi = \langle \Delta \hat{Q} \Delta \hat{P} + \Delta \hat{P} \Delta
\hat{Q} \rangle/2 \rho$ which, as is clear from its definition, reflects 
the correlation between $\Delta \hat{Q}$ and $\Delta \hat{P}$. 
Hamilton's equations for the extended open system are now written as 

\begin{eqnarray}
\frac{d  Q }{dt}  & = &  \frac{P}{M}  \;, \\
\frac{d  P }{dt}  &  = & 2(\mu-6\lambda \rho^2)Q-4 \lambda Q^3 \nonumber \\
& & +m \sum_{n=1}^{N} \omega_n^2(q_n-Q) \;, \\
\frac{d \rho}{dt}  & = & \frac{\Pi}{M}  \;, \\
\frac{d \Pi}{dt}  & = & \frac{\hbar^2}{4M\rho^3}+
2\left(\mu-6\lambda Q^2-\frac{m}{2}\sum_{n=1}^{N} \omega_n^2\right)\rho \nonumber \\
& & -12 \lambda \rho^3 \;, \\
\frac{d  q_n }{dt}  & = &  \frac{p_n}{m}  \;, \\
\frac{d  p_n }{dt}  &  = &  -m  \omega_n^2(q_n-Q) \;,
 \end{eqnarray}
corresponding to an extended Hamiltonian for the total system of the form

\begin{figure}[t] 
\begin{center}
{\includegraphics[width=\columnwidth]{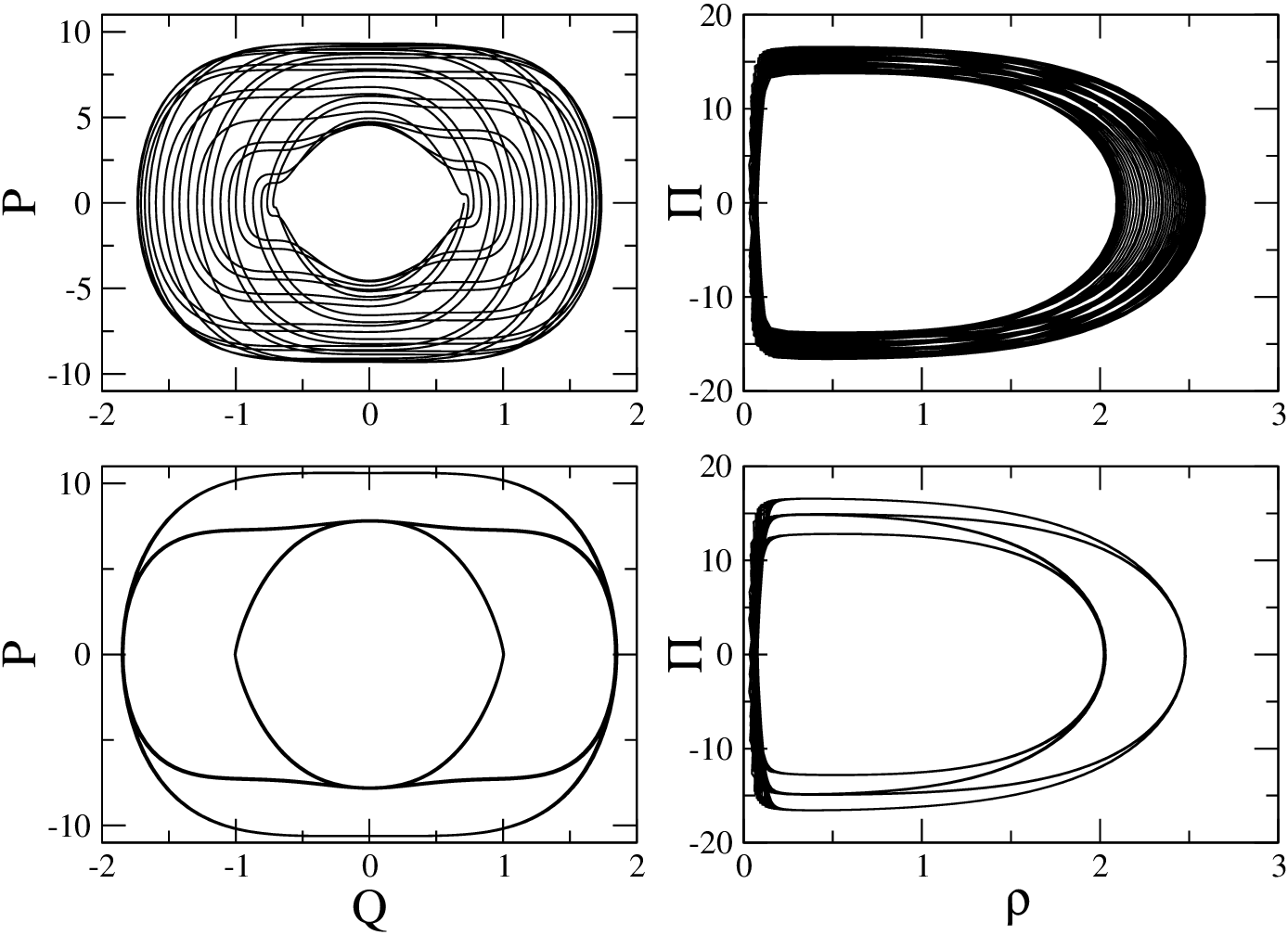}}
\caption{Test particle's phase space $(Q,P)$ and fluctuational phase space $(\rho,\Pi)$ for 
the case of $\mu=1, \lambda=1$, $M=1$. Initial conditions in $(Q,P,\rho,\Pi)$ space 
are $(1/\sqrt{2},0,0.03,0)$, corresponding to the test particle sitting 
in the minimum of the potential well with zero initial momentum (top plots), 
and  to $(1,0,0.03,0)$ for which the test particle starts away from the minimum of 
the double-well (bottom plots). The time evolution lasts for 5 periods where the cycle 
time is defined by the effective period of the intrawell oscillations, $\tau=2 \pi/\sqrt{2\mu}$. 
Even starting at the minimum of the potential well with zero kinetic energy in real phase 
space, the kinetic energy associated with fluctuations feeds back on the phase space 
giving rise to quantum tunneling.}  
\label{Josephsonfig2}
\end{center}
\end{figure}

\begin{eqnarray}
& & H(Q,P;\rho,\Pi;q_n,p_n) =  \frac{P^2}{2M}-\mu Q^2 +\lambda Q^4+6 \lambda Q^2 \rho^2 \nonumber \\
& & +\frac{1}{2}m \sum_{n=1}^{N}\omega_n^2 (q_n-Q)^2+\frac{\Pi^2}{2M}+\frac{\hbar^2}{8M \rho^2}
-\mu \rho^2 +3 \lambda \rho^4 \nonumber \\
& & +\frac{1}{2}m(\sum_{n=1}^{N} \omega_n^2)\rho^2+\sum_{n=1}^{N} \frac{p_n^2}{2m}.
\end{eqnarray}
Together, the $4+2N$ Hamilton equations fully describe the evolution of both the centroid 
and the spreading of the wave packet associated with the particle experiencing both the double-well 
potential and the classical motion of each of the particles making the heat bath, including 
the back-action effect of the particle on the heat bath as seen in Eq. (17). 
With respect to the classical Hamilton's equation, there is a renormalization term in the 
quadratic coefficient, and cross-terms relating fluctuational and configurational dynamics, 
written as $-12 \lambda \rho^2 Q$ and $-12 \lambda Q^2 \rho$, respectively. 
The presence of these cross-terms (absent in the case of a harmonic oscillator, 
which may be obtained as a specific case for $\mu <0$, $\lambda=0$) shows that 
configurational and fluctuation dynamics are reciprocally interrelated, at variance 
with the case of the harmonic oscillator.

\begin{figure}[t] 
\begin{center}
{\includegraphics[width=\columnwidth]{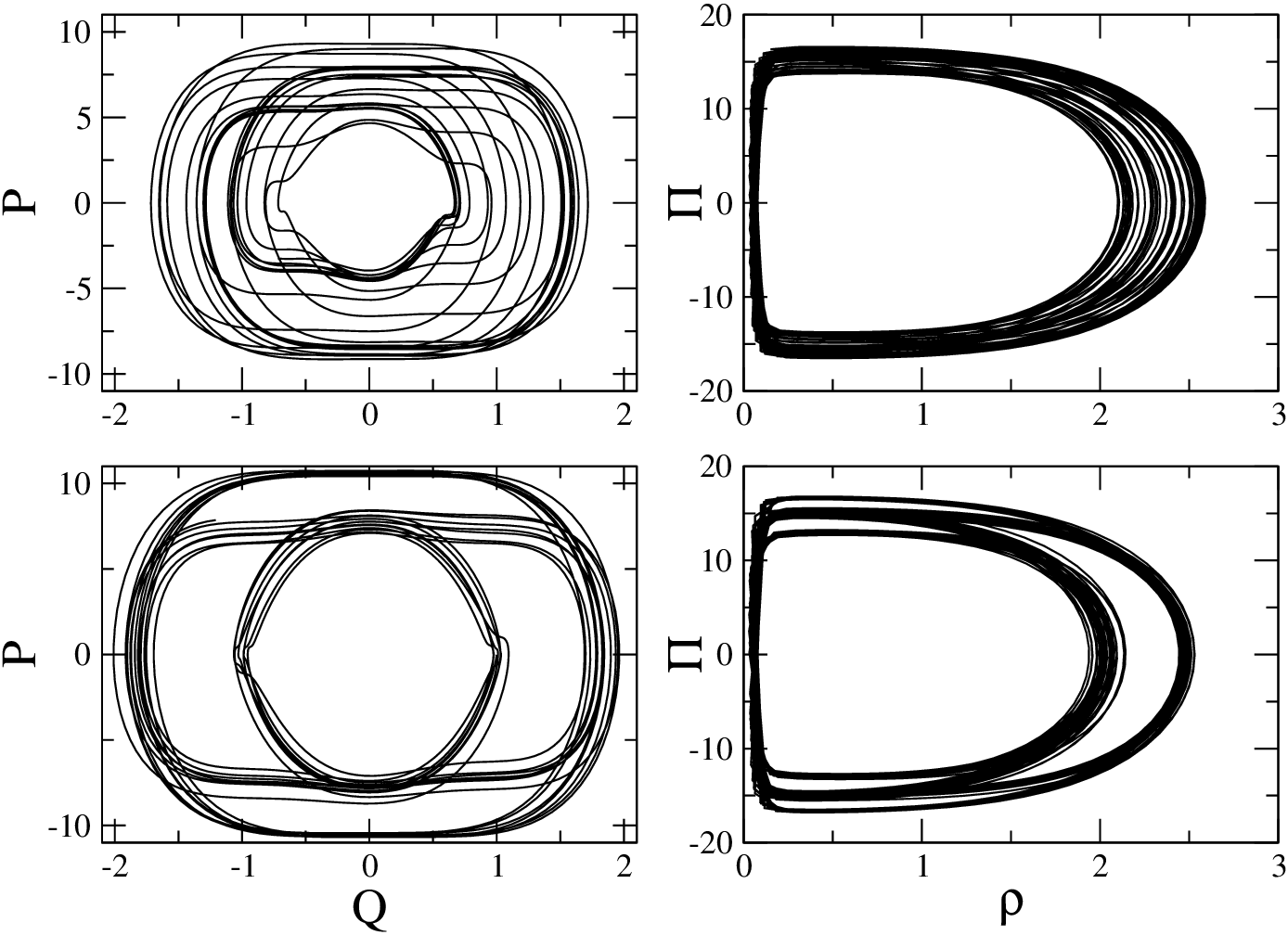}}
\caption{Same as previous figure but with the test particle in contact with a 
bath having a temperature $T=10^{-6}$.  The heat bath is made of 2000 oscillators 
with frequencies uniformly distributed in the range $[0,10]$, and mass $m=10^{-9}$.} 
\label{Josephsonfig3}
\end{center}
\end{figure}

\section{Numerical results}
We have numerically integrated the equations of motion under a wide range of bath and test particle
conditions using a variable-step, predictor corrector integration scheme, with absolute and relative 
errors in the $10^{-10}-10^{-12}$ range. In particular, we have explored the test particle dynamics for 
different initial conditions in the presence of the bath. The bath oscillators had energy drawn from 
a Boltzmann distribution and the number of oscillators, the mass of the harmonic oscillators, the angular 
frequency bounds as well as the spectral distribution between these two extrema were all varied. 
For the results shown we have chosen, for simplicity and for contrasting with the 
findings in~\cite{Taylor}, a uniform distribution of frequencies. 
However, more generic distributions \cite{LepriPhysRep} are easily analyzed with our numerical 
tools, although we do not expect any new or novel qualitative features to emerge~\cite{Wei}.

\begin{figure}[t] 
\begin{center}
{\includegraphics[width=\columnwidth]{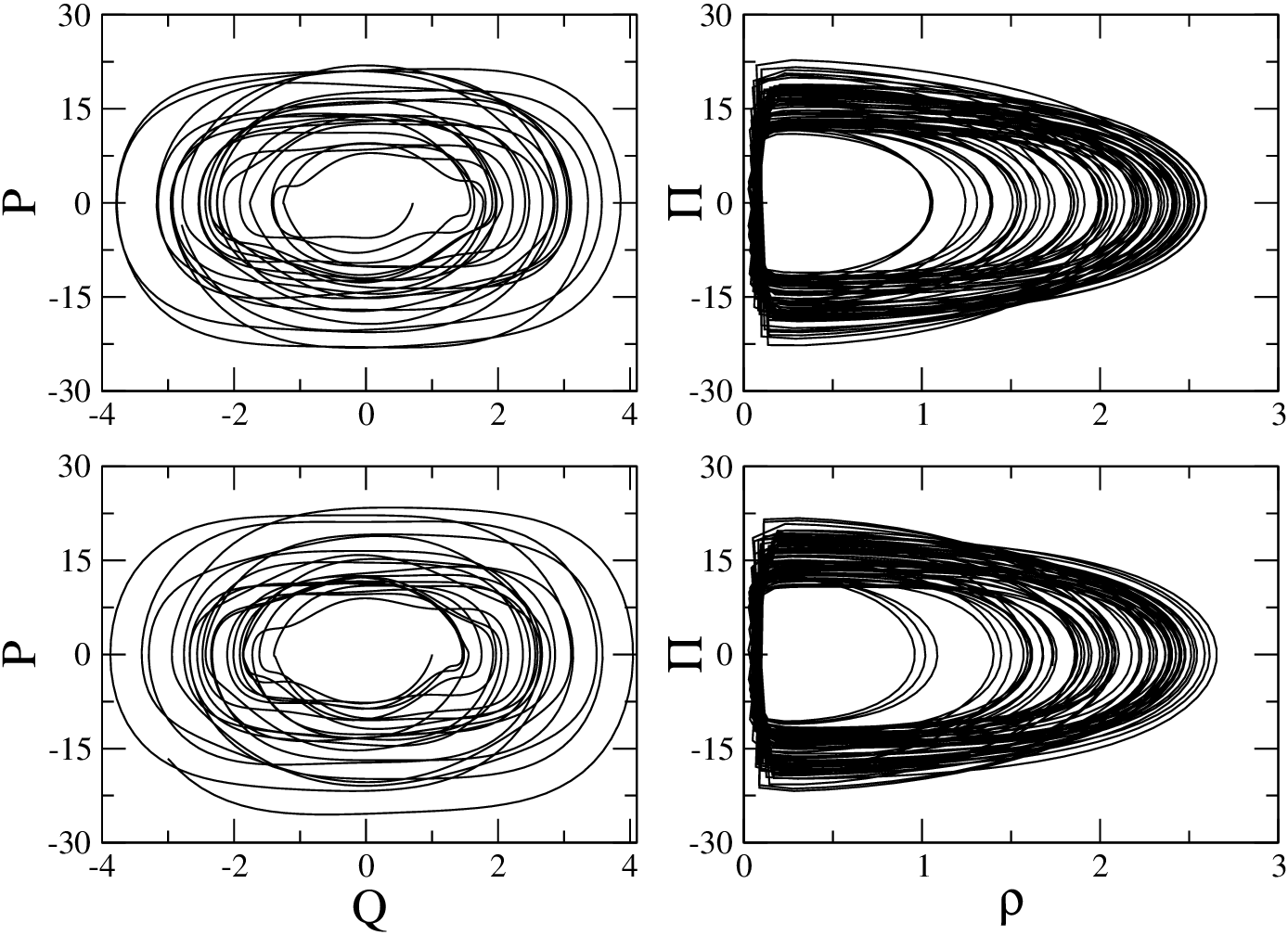}}
\caption{Same as previous figure but with the test particle in contact with a 
bath having a temperature $T=10^{-3}$, corresponding to thermally activated hopping processes.}
\label{Josephsonfig4}
\end{center}
\end{figure}

\begin{figure*}[t] 
\begin{center}
\includegraphics[width=\textwidth]{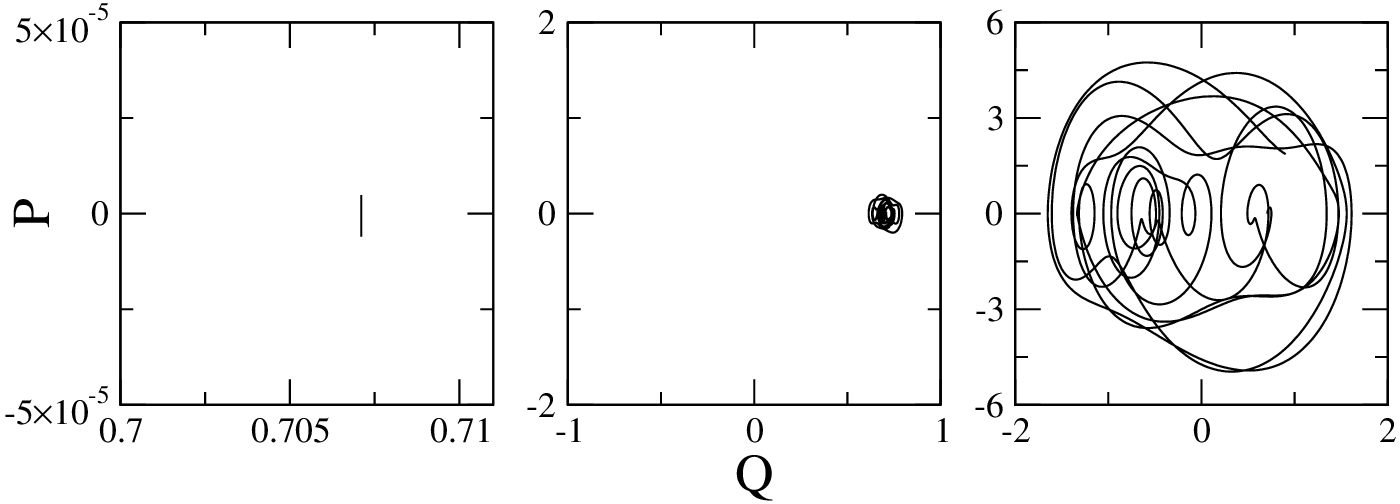}
\caption{Trajectories in phase space for the classical case ($\hbar=0, \rho=0$) for a 
particle in a double-well without external heat bath (left), and in contact with a heat 
bath at $T=10^{-6}$ (center), and $T=10^{-3}$ (right), with initial condition very close to  
the minimum of the potential, such that there is nearly no evolution in the first case, some 
random motion without hopping in the second case, and effective thermal activation in the third case.}
\label{Josephsonfig5}
\end{center}
\end{figure*}

We begin with the absence of a bath and consider a particularly intriguing choice of 
initial conditions corresponding to the test particle being located at rest at the 
minimum of the double-well potential. 
In this case we expect no dynamics in the classical limit. The corresponding extended space
evolution is shown in the upper panels of Fig.~\ref{Josephsonfig2} and is considerably different.
This results from the purely quantum evolution 
in the $(\rho,\Pi)$ space and can be thought of arising from auxiliary kinetic energy associated 
with quantum fluctuations and resulting from the enforcement of the Heisenberg principle. 
This emphasizes the crucial role of the $(\rho,\Pi)$ space in providing fluctuation
energy to the regular phase space. It is also evident that there are initial conditions in the 
former space for which no tunneling dynamics is achieved, {\it i.e.} in a region centered 
around $\rho=1,\Pi=0$.
As far as we know, there has been little discussion of this interpretation of quantum tunneling 
resulting from the energy available in pure quantum fluctuations, even if the system is 
nominally, on average, at rest at a classical minimum of the potential. 
If the particle starts from initial conditions corresponding to higher energy, the dynamics 
changes as seen in the lower panels of Fig.~\ref{Josephsonfig2}. 
The particle dynamics starting with the same initial conditions is qualitatively 
quite different when the bath oscillators are included in the dynamics. 
In Figs.~\ref{Josephsonfig3} and ~\ref{Josephsonfig4}, we show cases where the 
bath temperature is varied.

If the temperature is not very large, as in Fig.~\ref{Josephsonfig3}, the generalized phase space 
trajectories of the two different initial conditions are clearly distinct. By contrast, as seen in 
Fig.~\ref{Josephsonfig4}, coupling to a high-temperature bath gives rise to nearly indistinguishable 
trajectories in the generalized phase space. The interpretation is that Fig.~\ref{Josephsonfig3} results 
from the combined action of the fluctuational quantum energy and the thermal energy, while in 
Fig.~\ref{Josephsonfig4} the thermal activation processes dominate the dynamics. 
Also, the back-action in the fluctuation sector of phase space is more clearly manifest 
in Fig.~\ref{Josephsonfig4}, as seen from the wider excursions in $\Pi$ due to the stronger 
correlation to the configurational dynamics due to the cross-terms discussed after Eq. (8).
By way of contrast, we show in Fig.~\ref{Josephsonfig5} the strictly classical counterpart of 
the motion in the absence and presence of a heat bath with increasing temperature and 
the associated growth in thermal activation.

\begin{figure}[t] 
\begin{center}
{\includegraphics[width=\columnwidth]{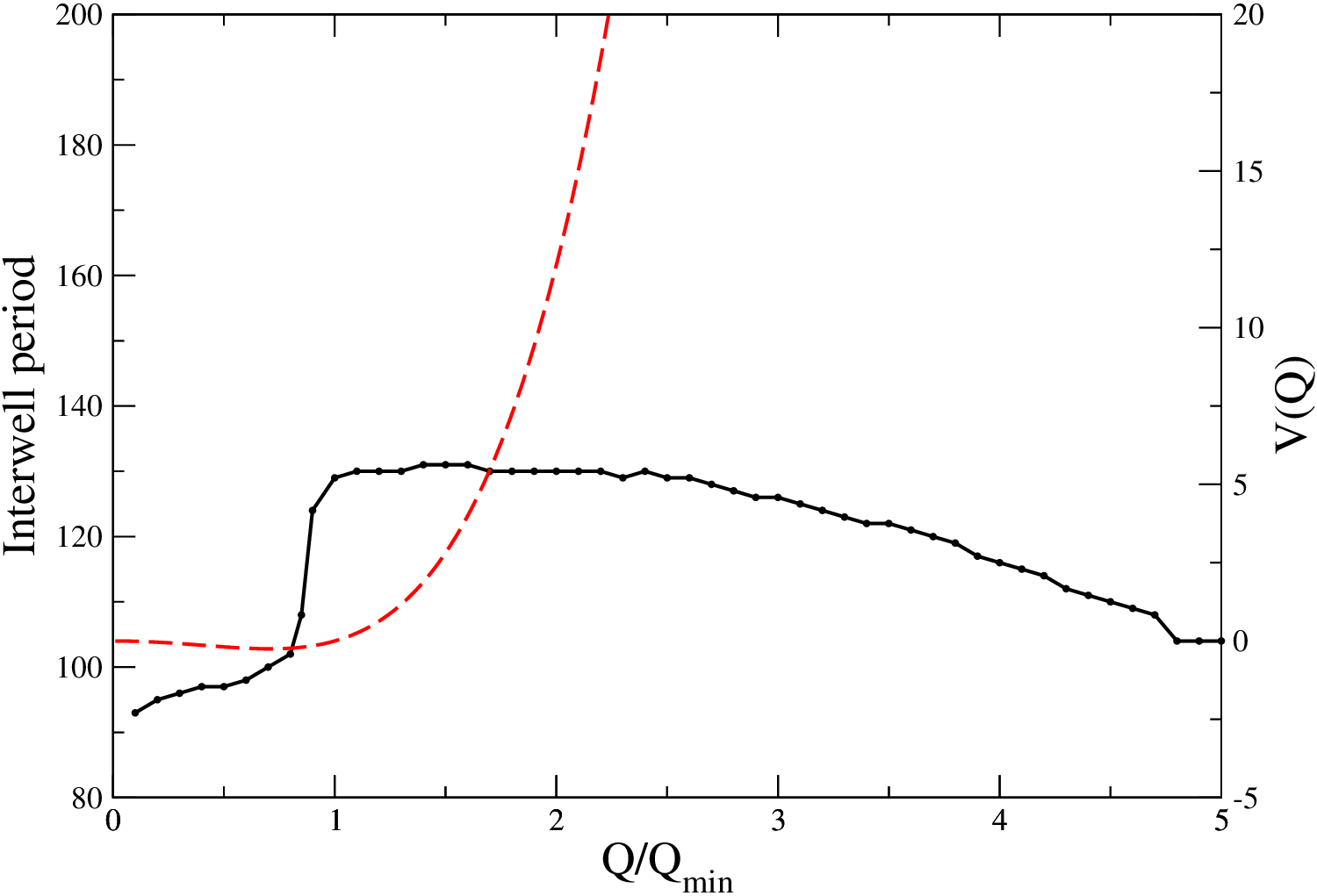}}
\caption{(Color online) Interwell period versus the initial position of the 
test particle (black continuous line, left vertical scale) and corresponding potential 
energy (red dashed line, right vertical scale). A dramatic increase in period corresponds 
to initial position at the minimum of the potential, with the interpretation 
of a lower transition probability. Here the temperature of the bath is assumed 
to be zero and $\mu=\lambda=1$, and the initial conditions for $P,\rho,\Pi$ are the
same as in Fig.~\ref{Josephsonfig2}.}
\label{Josephsonfig6}
\end{center}
\end{figure}

The dependence of the interwell period of oscillation upon the initial position of the test 
particle with a zero-temperature heat bath is shown in Fig.~\ref{Josephsonfig6}, where the 
interwell period grows quickly in a narrow region of $Q/Q_{\mathrm{min}}$ around unity. 
On closer inspection, it is evident that the maximum slope of the interwell period 
with respect to $Q/Q_{\mathrm{min}}$ occurs away from
the actual minimum of the potential, consistent with the fact that energy coming from the $(\rho,\Pi)$ 
space is also available for hopping. Moreover, the presence of the bath, even at 
zero-temperature, renormalizes the bare potential of Eq. (1), as discussed in 
Fig.~\ref{Josephsonfig1}, resulting in a minimum at a value smaller than $Q/Q_{\mathrm{min}}=1$.
The increase in period is a manifestation of a lower transition probability, and one may extend 
this inference to baths at different temperatures. 
Thus, the crossover regime between hopping due to individual or collective mechanisms 
of quantum and thermal fluctuations can be explored. 
However, a closer inspection of the time dependence of $Q$ with an FFT analysis shows that, 
even for a closed system, there are various frequencies contributing to the motion. 

\begin{figure}[t]
\begin{center}
{\includegraphics[width=\columnwidth]{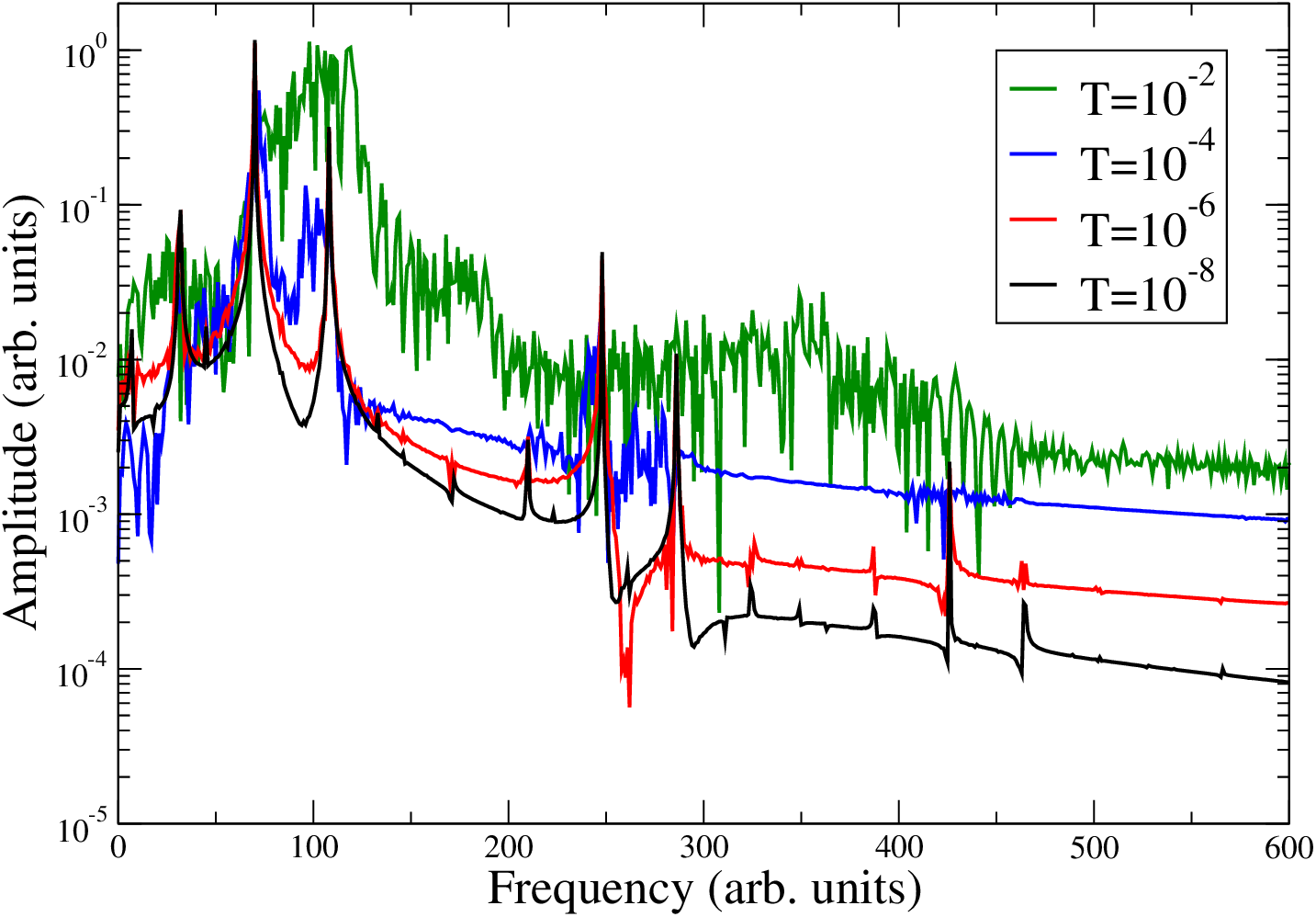}}
\caption{(Color online) Fast Fourier transform of the time-dependent position of 
the particle in the double-well potential for different temperatures of the reservoir, 
from top to bottom curve on the right side of the plot, T=$10^{-2}$ (green curve), T=$10^{-4}$ 
(blue), T=$10^{-6}$ (red), and T=$10^{-8}$ (black), in arbitrary units. 
The regime of thermal activation is marked by the presence of a broad spectrum of frequencies, 
while at low temperatures the motion has a small set of dominant frequencies, peaked around 
curves progressively broadened as the temperature is increased.
The motion is sampled for $2^{13}=8192$ time steps, with initial conditions and 
the same system parameters as in the bottom plots of Fig.~\ref{Josephsonfig2}, 
and 2,000 oscillators in the heat bath.}
\label{Josephsonfig7}
\end{center}
\end{figure} 

Therefore the interwell period of oscillation has to be read as a coarse-grained
observable associated with the Fourier spectrum of $Q(t)$. The FFT spectrum is a more 
detailed indicator and strongly depends on the temperature of the heat bath, as shown 
in Fig.~\ref{Josephsonfig7}. At low temperature there are few prominent peaks, while 
at the highest explored temperature the peaks are barely recognizable as they are 
submerged in a broadband spectrum of frequencies. 

This suggests that precision thermometry of tunneling electrons should be achieved by 
looking at the FFT spectrum of the Josephson current, or related quantities such as 
the integral of the FFT spectrum over the frequency range or the frequency of the peak 
value in the FFT spectrum, which is manifestly broadened as the temperature increases. 
Thermometry based on FFT analysis of transport properties could be more robust than the 
one based on decoherence of the interference fringes, and could be implemented in the same 
measurement run by nondestructive imaging techniques in the case of ultracold atoms \cite{Andrews}. 
While thermometry using Josephson junctions is already available in the solid-state~\cite{Faivre}, 
thermometry of ultracold atoms is a crucial issue~\cite{DeMarco} and the influence of a finite 
temperature environment on quantum tunneling in a double-well trap~\cite{Ananikian,Gottlieb} has 
been observed~\cite{Gati,Eckel}. 

\section{Conclusions}

In conclusion, we have discussed within a unified framework the interplay between quantum 
and classical fluctuations in the nontrivial case of a open double-well system 
using the Ehrenfest approach implemented through the Gaussian approximation.
We have provided plausibility arguments according to which this approximation 
should become increasingly reliable as the heat bath dominates the dynamics, which 
happens in the limits of high temperature and/or large number of bath oscillators. 
At finite times, the method may include the case of resetting dynamics like the one 
of superconducting electronics, or possibly time-dependent situations such as the ones 
related to stochastic resonance~\cite{Gammaitoni}, readily achievable by adding a time-dependent 
tilting force term to the double-well potential. From the Ehrenfest viewpoint, stochastic resonance 
appears as a specific example of control of the test-particle Hamiltonian parameters in order
to maximize performance, as measured by the signal-to-noise ratio when the test particle is 
subjected to an external perturbation  and the fluctuations induced by the heat bath.

The inclusion of the external environment shows also that the Ehrenfest approximation is 
expected to be more reliable, or better to be able to approximate the quantum dynamics for a 
longer time, as we increase the coupling to the environment by increasing the heat bath temperature. 
This is not surprising as in this limit the dynamics of the double-well is increasingly ruled 
by the (classical) heat bath. This is also suggestive of a possible solution to the controversial 
issue of the validity of the Ehrenfest approximation to systems other than harmonic ones. 
As neatly pointed out in~\cite{Habib}, {\it when closed-system classical and quantum dynamics are 
treated in Gaussian approximation, they are in fact identical}, which simplifies the situation as 
any difference in their average behavior is then due to the breakdown of the Ehrenfest approximation. 
Therefore in the case of an {\it open} system the situation is more forgiving, and the breakdown of 
the Ehrenfest approximation occurs either on much longer timescales, of order of $\hbar/\Delta E$ 
(where $\Delta E$ is the minimum spacing of the relevant energies significantly contributing 
to the motion) or never for conditions under which the effective potential becomes monostable. 
As seen in Fig.~\ref{Josephsonfig7} for higher temperatures, an increasing amount of 
harmonics contribute therefore making $\Delta E$ smaller with respect to the closed system. 
This should alleviate both the issues raised for a closed system in \cite{Hasegawa1,Hasegawa2} 
and the need for using a superposition of Gaussian wavepackets suggested in \cite{Zoppe}.

Less trivial is the dependence of the dynamics on $N$ in the macroscopic limit, $N\rightarrow \infty$ . 
Based on the considerations on the phase transition occurring when the number of oscillators exceed
$N_{\mathrm{crit}}$ as seen in Eq. (4), we notice that the same effect is applicable to Eqs. (13) and (15), 
where the cross-term $6 \lambda Q^2 \rho^2$ no longer affects the qualitative analysis. 
This means that for the number of oscillators $N > N_{\mathrm{crit}}$ the effective potentials in both $Q$ and 
$\rho$ are monostable. Thus, the dynamical behavior will be qualitatively quite similar to the 
case of a harmonic oscillator for which the Ehrenfest dynamics is valid for all times.  
Since typically the number of degrees of freedom of a bath also depends on temperature, 
with more degrees of freedom becoming unfrozen with increasing temperatures, there is a range
of possible behavior in this macroscopic limit, depending on the concrete model schematizing the heat bath. We also note that the Caldeira-Leggett model, originally introduced to describe the influence of a high-temperature, infinite-bandwidth bath on a double-well system, is expected to fail at low 
temperatures for which ultraviolet cut-offs in the frequency spectrum of the bath oscillators are 
necessarily operative. This expected failure is consistent with our finding of a phase transition 
to a monostable potential for the case of a finite bandwidth and a number of oscillators in the 
bath above a well-defined threshold, a situation which could be indeed realized in low-temperature setups.

From the experimental perspective, in the framework of Bose condensates the system can 
be further generalized by including interatomic interactions~\cite{Pitaevskii,LeggettRMP}. 
For instance, in the experiment~\cite{Gati}, the double-well is created from a pre-existing 
harmonic potential via a slow ramping up of a barrier. It remains to be understood if 
residual non-adiabaticity in the ramp-up may be a cause for temperature increase of the sample. 
The formalism developed here allows the implementation of strategies, such as frictionless 
cooling~\cite{Yuce,Chen,Torrontegui,Torrontegui1,Torronteguirev}, capable of minimizing 
this heating source.  Work on extending the Caldeira-Leggett model in the framework of 
trapped atoms is ongoing~\cite{OnBa}. However, the functional dependence of the interacting 
Hamiltonian, necessary for making the test-particle bath-particle interaction local, precludes 
closure of the system of equations even in the Gaussian approximation, so all current 
indications are that this formalism may not be trivially extended to this important 
class of many-body systems.


\begin{thebibliography}{99}

\bibitem{Ehrenfest} P. Ehrenfest, Zeit. f. Physik \textbf{45}, 455 (1927).

\bibitem{Chaos} See, for example, 
M. D. Feit and J. A. Fleck, J. Chem. Phys. \textbf{80}, 2578 (1984); 
S. Tomsovic and E. J. Heller, \prl \textbf{67}, 664 (1991); 
S. Habib, K. Shizume, and W. H. Zurek, \prl \textbf{80}, 4361 (1998);
P. G. Silvestrov and C. W. J. Beenakker, \pre \textbf{65}, 035208 (2002).

\bibitem{PS1} A. K. Pattanayak and W. C. Schieve, Phys. Rev. A \textbf{46}, 1821 (1992);  

\bibitem{PS2} A. K. Pattanayak and W. C. Schieve, Phys. Rev. Lett. \textbf{72}, 2855 (1994).

\bibitem{PS3} A. K. Pattanayak and W. C. Schieve, Phys. Rev. E \textbf{50}, 3601 (1994).

\bibitem{SM} B. Sundaram and P. W. Milonni, \pre \textbf{51}, 1971 (1995).

\bibitem{Gammaitoni} L. Gammaitoni, P. H\"anggi, P. Jung, and F. Marchesoni, 
Rev. Mod. Phys. \textbf{70}, 223 (1998).

\bibitem{Caldeira} A. O. Caldeira and A. J. Leggett, Phys. Rev. Lett. \textbf{46}, 211 (1981).

\bibitem{CaldeiraRMP} A. J. Leggett, S. Chakravarty, A. T. Dorsey, M. P. A. Fisher, A. Garg, and W. Zwerger, 
Rev. Mod. Phys. \textbf{59}, 1 (1987).

\bibitem{Jona} G. Jona-Lasinio, C. Presilla, and C. Toninelli, Phys. Rev. Lett. \textbf{88}, 123001 (2002). 

\bibitem{Choionba} S. Choi, R. Onofrio, and B. Sundaram, Phys. Rev. A \textbf{88}, 053401 (2013).

\bibitem{Taylor} S. T. Smith and R. Onofrio, Eur. Phys. J B \textbf{61}, 271 (2008).

\bibitem{LepriPhysRep} S. Lepri, R. Livi, and A. Politi, Phys. Rep. \textbf{377}, 1 (2003).

\bibitem{Wei} Q. Wei, S. T. Smith, and R. Onofrio, Phys. Rev. E \textbf{79}, 031128 (2009).

\bibitem{Andrews} M. R. Andrews, M.-O. Mewes, N. J. van Druten, D. S. Durfee, D. M. Kurn, 
and W. Ketterle, Science \textbf{273}, 84 (1996).

\bibitem{Faivre} T. Faivre, D. Golubev, and J. P. Pekola, Journ. Appl. Phys. \textbf{116}, 094302 (2014).

\bibitem{DeMarco} D. C. McKay and B. DeMarco, Rep. Prog. Phys. \textbf{74}, 054401 (2011).

\bibitem{Ananikian} D. Ananikian and T. Bergeman, Phys. Rev. A \textbf{73}, 013604 (2006).

\bibitem{Gottlieb} A. D. Gottlieb and T. Schumm, Phys. Rev. A \textbf{79}, 063601 (2009). 

\bibitem{Gati} R. Gati, B. Hemmerling, J. F\"olling, M. Albiez, and M. K. Oberthaler,
Phys. Rev. Lett. \textbf{96}, 130404 (2006).

\bibitem{Eckel} S. Eckel, J. G. Lee, F. Jendrzejewski, N. Murray, C. W. Clark, C. J. Lobb, 
W. D. Phillips, M. Edwards, and G. K. Campbell, Nature \textbf{506}, 200 (2014).

\bibitem{Habib} S. Habib, quant-ph/0406601 (1 Jun 2004).

\bibitem{Hasegawa1} H. Hasegawa, Phys. Rev. E \textbf{86}, 061104 (2012). 

\bibitem{Hasegawa2} H. Hasegawa, Phys. Lett. A \textbf{378}, 691 (2014).

\bibitem{Zoppe} J. O. Zoppe, M. L. Parkinson, and M. Messina, Chem. Phys. Lett. \textbf{407}, 308 (2005).

\bibitem{Pitaevskii} L. Pitaevskii and S. Stringari, Phys. Rev. Lett. \textbf{87}, 180402 (2001).

\bibitem{LeggettRMP} A. J. Leggett, Rev. Mod. Phys. \textbf{73}, 307 (2001).

\bibitem{Yuce} C. Yuce, A. Kilic, and A. Coruh, Phys. Scr. \textbf{74}, 114 (2006).

\bibitem{Chen} X. Chen, A. Ruschhaupt, S. Schmidt, A. del Campo, D. Gu\'ery-Odelin, and J. G. Muga, 
Phys. Rev. Lett. \textbf{104}, 063002 (2010).

\bibitem{Torrontegui} E. Torrontegui, S. Ib\'a\~nez, X. Chen, A. Ruschhaupt, 
D. Gu\'ery-Odelin, and J. G. Muga, Phys. Rev. A \textbf{83}, 013415 (2011).

\bibitem{Torrontegui1} E. Torrontegui, X. Chen, M. Modugno, A. Ruschhaupt, D. Gu\'ery-Odelin, 
and J. G. Muga, Phys. Rev. A 85, 033605 (2012).

\bibitem{Torronteguirev} E. Torrontegui, S. Ib\'a\~nez, S. Mart\'{i}nez-Garaot, M. Modugno, A. del Campo, 
D. Gu\'ery-Odelin, A. Ruschhaupt, X. Chen, and J. G. Muga, Adv. At. Mol. Opt. Phys. \textbf{62}, 117 (2013).

\bibitem{OnBa} R. Onofrio and B. Sundaram, Phys. Rev. A \textbf{92}, 033422 (2015).

\end{thebibliography}
\end{document}